\documentclass[useAMS,usenatbib,usegraphicx,footnotetext]{mn2e}

\usepackage{multirow}

\newcommand{\hii}{H\textsc{ii}}
\def\ks{km s$^{-1}$}

\def\m{$^\prime$}
\def\s{$^{\prime\prime}$}

\def\ss{$^{\mathrm s}$}
\def\cm3{cm$^{-3}$}

\def\2{$^{12}$CO}
\def\3{$^{13}$CO}
\def\H{HCO$^{+}$}
\def\msun{M$_\odot$}

\title[A view of some LMC HII regions through the 345 GHz window]{A view of Large Magellanic Cloud HII regions N159, N132, and N166 through the 345 GHz window}

\author[S. Paron et al.]{S. Paron$^{1,2}$\thanks{sparon@iafe.uba.ar},  M. E. Ortega$^{1}$, C. Fari\~{n}a$^{3}$, M. Cunningham$^{4}$, P. A. Jones$^{4}$, and M. Rubio$^{5}$\\ 
$^{1}$ Instituto de Astronom\'{\i}a y F\'{\i}sica del Espacio (IAFE),
             CC 67, Suc. 28, 1428 Buenos Aires, Argentina \\
$^{2}$ FADU and CBC, Universidad de Buenos Aires, Argentina \\ 
$^{3}$ Isaac Newton Group of Telescopes, E38700, La Palma, Spain\\
$^{4}$ School of Physics, University of New South Wales, Sydney, NSW 2052, Australia  \\
$^{5}$ Departamento de Astronom\'{\i}a, Universidad de Chile, Casilla 36-D, Santiago, Chile \\ 
}

\begin{document}

\date{Accepted XXXX. Received XXXX; in original form XXXX}

\pagerange{\pageref{firstpage}--\pageref{lastpage}} \pubyear{2015}

\maketitle

\label{firstpage}

\begin{abstract}
We present results obtained towards the \hii~regions N159, N166, and N132 from the emission of several molecular
lines in the 345 GHz window. 
Using ASTE we mapped a 2\farcm4 $\times$ 2\farcm4  region towards the
molecular cloud N159-W in the \3 J=3--2 line and observed several molecular lines at an IR peak very 
close to a massive young stellar object. \2 and \3 J=3--2 were observed towards two positions in N166 and one position in N132.
The \3 J=3--2 map of the N159-W cloud shows that the molecular peak is shifted southwest compared to the peak of the IR emission.
Towards the IR peak we detected emission 
from HCN, HNC, \H, C$_{2}$H J=4--3, CS J=7--6, and tentatively C$^{18}$O J=3--2. This is the first reported detection of 
these molecular lines in N159-W. The analysis of the C$_{2}$H line yields more evidence supporting that the chemistry 
involving this molecular species in compact and/or UC\hii~regions in the LMC should be similar
to that in Galactic ones. 
A non-LTE study of the CO emission suggests the presence of both cool and warm gas in the analysed region.
The same analysis for the CS, \H, HCN, and HNC shows that it is very likely that their emissions arise mainly from warm gas with
a density between $5 \times 10^{5}$ to some $10^{6}$ cm$^{-3}$.
The obtained $\frac{\rm HCN}{\rm HNC}$ abundance ratio greater than 1 is compatible with warm gas and with an star-forming scenario.
From the analysis of the molecular lines observed towards N132 and N166 we propose that both regions 
should have similar physical conditions, with densities of about 10$^{3}$ cm$^{-3}$.

\end{abstract}

\begin{keywords}
galaxies: ISM -- (galaxies:) Magellanic Clouds -- (ISM:) \hii~regions -- ISM: molecules
\end{keywords}

\section{Introduction}

The Large Magellanic Cloud (LMC), at only 49.97 kpc \citep{pietr13} is a gas rich environment with 
reduced metallicity (Z about half of the Galactic value) that allow us to obtain detailed observational data to study the physical 
processes leading to massive star formation in an interstellar medium (ISM) that may be comparable, to certain degree, to those 
conditions of some star-forming sites in our Galaxy in the early stages (see for example \citealt{yamada13}). 
Several global surveys of the molecular component in the LMC have been done so far, mainly in the CO J=1--0 emission,
in increasing resolutions, starting with an angular resolution of $\sim$10\m~\citep{cohen88,rubio91}, to the latest made with 
NANTEN with 2\farcm6 of angular resolution \citep{fukui08}. These surveys were complemented with observations at better angular resolution of 
known individual cloud complexes, such as the ESO SEST Key Programme (e.g. \citealt{isra93,garay02,isra03}), the Magellanic Mopra Assessment 
(e.g. \citealt{wong11}), and the second survey of molecular clouds with NANTEN \citep{kawa09}.

\begin{figure*}
\centering
\includegraphics[width=15cm]{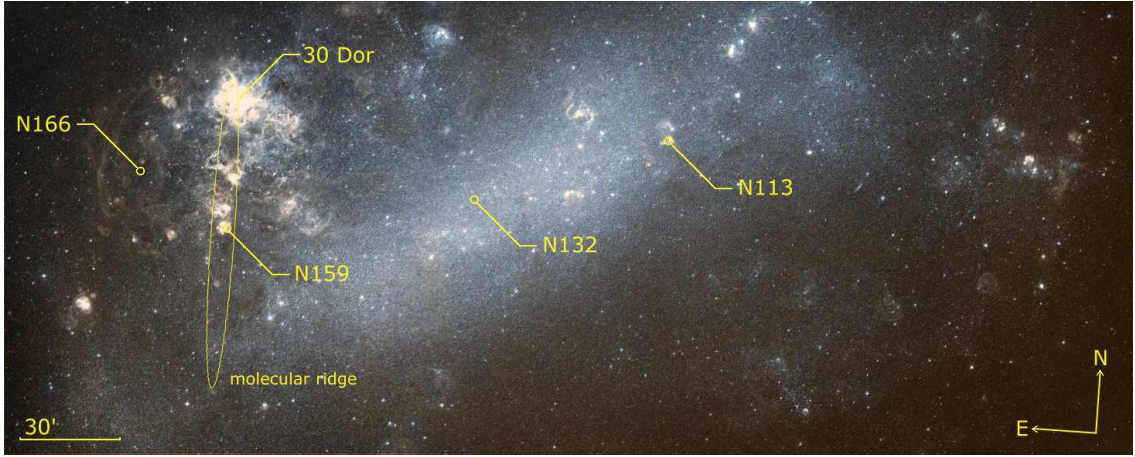}
\caption{Main area of the LMC in which the location of the \hii~regions studied in this paper (N159, N132, and N166) and N133 from 
\citet{paron14} are indicated. As reference, 30 Dor and the location of the LMC molecular ridge are also showed. The background is 
an optical image from DSS  Hierarchical Progressive Survey retrieved via Aladin \citep{bonna00}.}
\label{global}
\end{figure*}

Most of the molecular line surveys and individual observations towards the LMC do not cover the 345 GHz window, which contains
several molecular lines that provide substantial information about the physical and chemical conditions.  
With the idea of carry out a comparative study of the physical conditions of the molecular gas in \hii~regions 
in the LMC using molecular lines in the 345 GHz windows, we have made observations with the 
Atacama Submillimetre Telescope Experiment (ASTE) towards different regions. In \citet{paron14} (hereafter Paper\,I), 
we published the results from N113 study. In this paper we add the results from the study of other three \hii~regions: N159, N132, and N166. 
Although these regions share the same 
global characteristics in terms of metallicity, they are located in a completely different environment within the LMC (see Figure \ref{global}).

N159 is by far the most studied \hii~region of the set. Situated approximately 600 pc in projection south of 30 Dor, in the CO molecular 
ridge, it is a region likely perturbed by the interaction with the Milky Way halo \citep{ott08}. 
The N159 complex was classified, in the NANTEN catalog compiled by \citet{fukui08}, as a type III giant molecular cloud (GMC), that is
a GMC with \hii~regions and young star clusters. This complex is populated by young 
massive stars (e.g. \citealt{fari09}) and presents numerous features characteristic of active star formation regions. \citet{gatley81} 
discovered the first extragalactic protostar here, and \citet{cas81} the first Type I extragalactic OH maser. 
It is known that N159 hosts massive embedded young stellar objects (YSOs), a maser source, and several ultracompact \hii~regions \citep{chen10}. 
The carbon in the gaseous phase of the whole complex was studied in detail by \citet{bol08}, whereas \citet{mizuno10} 
studied the warm dense molecular gas. Recently \citet{fukui15}, using ALMA \3 J=2--1 observations,
discovered the first extragalactic protostellar molecular outflows towards this region. 

N166 located in projection about 550 pc south-east of 30 Dor, between 30 Dor and N159 to the east of the CO molecular ridge. This region, 
associated with the molecular cloud DEM 310 \citep{davies76} and the giant molecular Complex-37 \citep{garay02}, 
was cataloged as type II GMC (a GMC with \hii~regions only) in the NANTEN catalog. 
\citet{mina08} studied the \2 J=1--0 and J=3--2 emission towards five clumps in N166, and suggest that this 
region is in a younger phase of star formation than N159 as density 
has not yet reached high enough to start the born of massive stars. 

N132 located in projection about 1200 pc south-west of 30 Dor, on the northern edge of LMC bar, is associated with the molecular clouds 
DEM 172, 173 and 186 \citep{davies76,kawa09}. As in the case of N166 this region is also a GMC Type II.
This region has not been particularly studied apart of a global characterization in which the H$_{2}$ column density is estimated 
and the H$_{2}$--CO ratio determined \citep{isra97}.

In this paper we present the study we have carried out with new observations made towards the LMC \hii~regions N159, N132, and N166 
in a set of molecular lines in the 345 GHz window: \2 and \3 J=3--2 and the unexplored lines (in N159) HCN, HNC, \H, 
and C$_{2}$H (in the J=4--3 transition), CS J=7--6, and C$^{18}$O J=3--2.

\section{Observations and data reduction}

The molecular observations were performed between July and August 2010 with the 10 m ASTE telescope \citep{ezawa04}. 
The CATS345 GHz band receiver, a two-single band SIS receiver remotely tunable in the LO frequency range of 324-372 GHz, was used. 
The XF digital spectrometer was set to a bandwidth and spectral resolution of 128 MHz and 125 kHz, respectively.
The spectral velocity resolution was 0.11 \ks~and the half-power beamwidth (HPBW) was 22\s~at 345 GHz. The system temperature
varied from T$_{\rm sys} = 150$ to 250 K and the main-beam efficiency was $\eta_{\rm mb} \sim 0.65$.
The conversion factor to convert from Kelvin to Jansky is 78.3 (from T$_{\rm A}$).

The data were reduced with NEWSTAR\footnote{Reduction software based on AIPS developed at NRAO,
extended to treat single dish data with a graphical user interface (GUI).} and the spectra processed using the XSpec software 
package\footnote{XSpec is a spectral line reduction package for astronomy which has been developed by Per Bergman at Onsala Space Observatory.}. 
The spectra were Hanning smoothed to improve the signal-to-noise ratio, and in some cases, a boxcar smoothing was also applied.
Polynomials between first and third order were used for baseline fitting.

\begin{table}
\caption{Observed positions.}
\label{positions}
\begin{tabular}{lcc}
\hline
\hline
Region       &   R.A. (J2000)      & Dec. (J2000)      \\
\hline
N159         &  5:39:37.1          & $-$69:45:24.9     \\
N132         &  5:25:04.1          & $-$69:40:43.6     \\
N166-A   &  5:44:23.9          & $-$69:26:22.6     \\
N166-B   &  5:44:36.0          & $-$69:25:28.5     \\     
\hline
\end{tabular}
\end{table}

Several molecular lines in the 345 GHz window were observed towards the regions N159, N132, and N166.
The observed positions are indicated in Table\,\ref{positions}.  In
the case of N166 two different positions were observed, which are indicated
as A and B in the Table. These observations were performed in position
switching mode.
Additionally, we mapped a 2\farcm4 $\times$ 2\farcm4
region towards N159 centred at R.A.$=$5:39:38.3, dec.$=-$69:45:19.6 (J2000) in the \3 J=3--2 line.
This observation was performed in on-the-fly mapping mode achieving an angular sampling of 6\s.

\begin{figure}
\centering
\includegraphics[width=8.5cm]{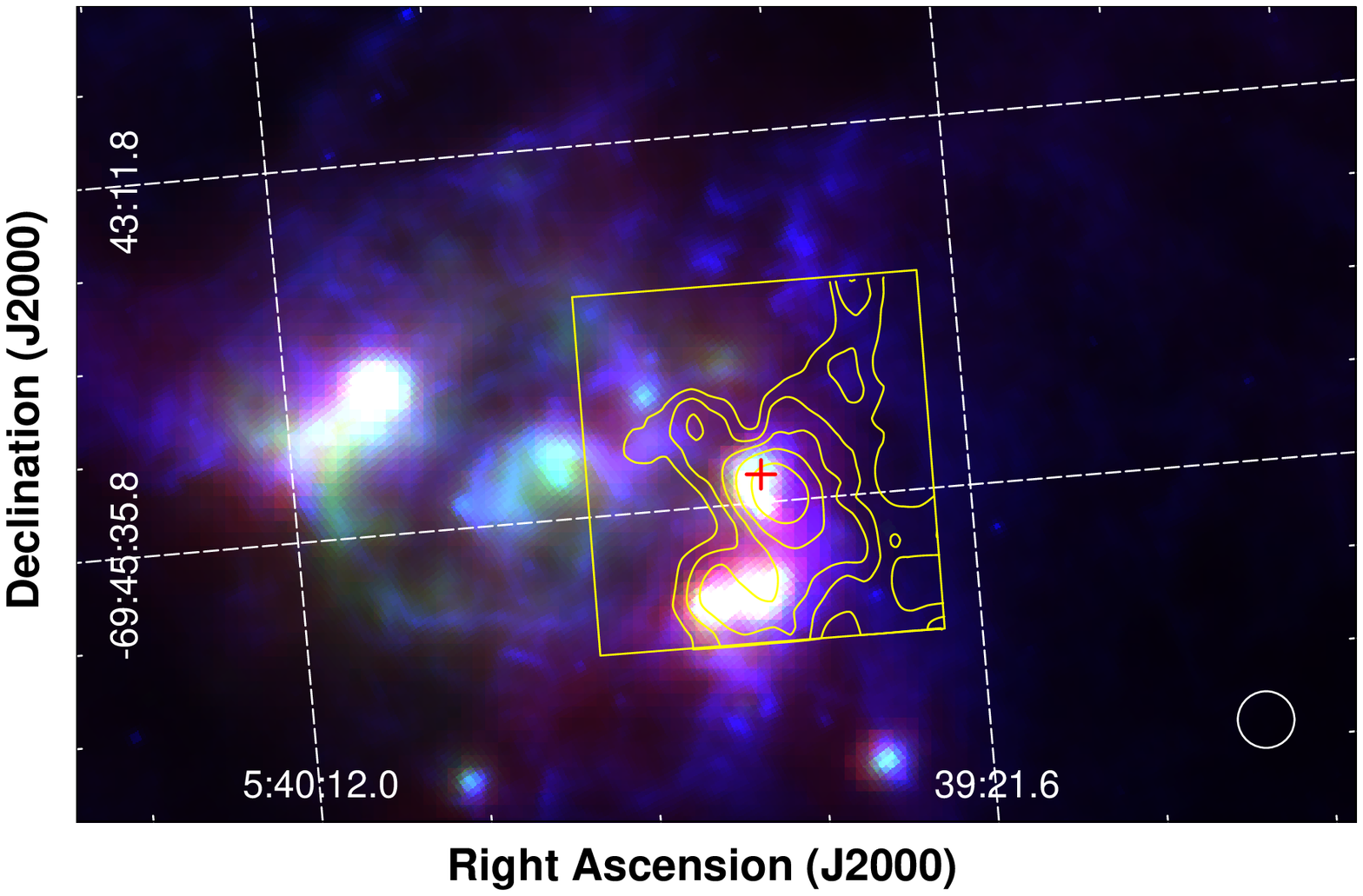}
\caption{Three-colour image of N159 where the 8, 24, and 70 $\mu$m emission toward N159 obtained from the IRAC and MIPS cameras of the Spitzer
Space Telescope (from SAGE Spitzer; \citealt{meixner06})  are presented in blue, green, and red, respectively.
The yellow box shows the region mapped in the \3 J=3--2 line with
an angular resolution of 22\s. The contours correspond to the \3 J=3--2 emission integrated between 225 and 250 \ks~with
levels of 3, 5, 7, 10, and 15 K \ks. The rms noise is 0.4 K \ks.
The red cross indicates the position where single spectra of several molecular lines were observed.
The beam size of the observations is included at the bottom right corner.}
\label{n159map}
\end{figure}

\begin{figure*}
\centering
\includegraphics[width=18cm]{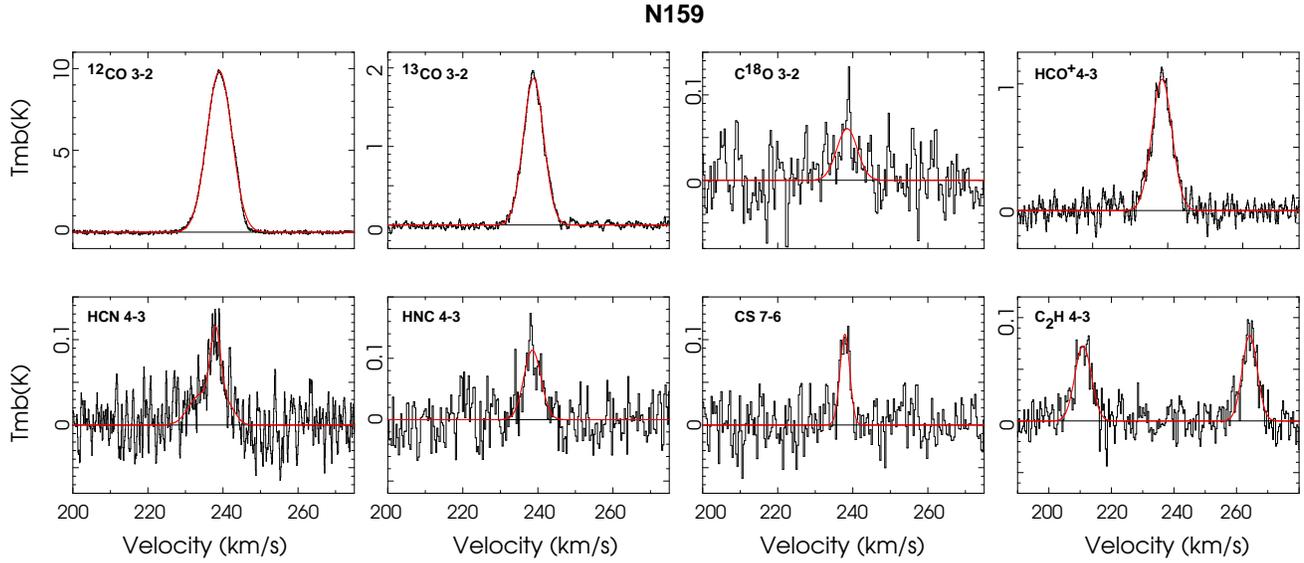}
\caption{Spectra of the detected molecular lines towards N159-W. The Gaussian fitting to each
spectrum is shown in red.}
\label{spectraN159}
\end{figure*}

\section{Results}

Figure\,\ref{n159map} is a three-colour image displaying the mid/far-IR emission in the N159 area where the mapped
region in the \3 J=3--2 line is indicated with a yellow square. The \3 J=3--2 emission integrated
between 225 and 250 \ks~is presented in contours. The surveyed region corresponds to the molecular cloud N159-W
\citep{joha98,bolatto00} which hosts several massive young stellar objects (MYSOs) \citep{chen10}. 
Moreover, recently \citet{fukui15}, using ALMA \3 J=2--1 observations,  
discovered the first extragalactic protostellar molecular outflows towards this region.

From the \3 J=3--2 map of N59-W and by assuming local thermodynamic equilibrium (LTE) we roughly estimate the molecular mass following the
same procedure as done in Paper\,I. From the \2 J=3--2 peak temperature (see Table\,\ref{lines})
we derive an excitation temperature of T$_{\rm ex} \sim 17$ K.  Using the peak \2 and \3 temperatures ratio,
we obtain the optical depths $\tau_{12} \sim 10.5$ and $\tau_{12} \sim 0.2$, which shows that the \3
J=3--2 line is optically thin. Once obtained the \3 column density (see Equation\,3 in Paper\,I),
we assumed an abundance ratio of [\3/H$_{2}$] $= 5.8 \times 10^{-7}$ \citep{heik99} to derive the H$_{2}$ column density.
Finally, the molecular mass was estimated from:
$${\rm M} = \mu~m_{{\rm H}} \sum_{i}{\left[ D^{2}~\Omega_{i}~{\rm N_{\it i}(H_{2}}) \right]},$$ 
where $\Omega$ is the solid angle subtended by the beam size, $D$ is the distance (50~kpc), $m_{\rm H}$ is the hydrogen mass,
and $\mu$ is the mean molecular weight, assumed to be 2.8 by taking into account a relative helium abundance of 25 \%.
The summation was performed over all the beam positions belonging to the molecular
structure delimited by the 7 K \ks~contour displayed in Fig.\,\ref{n159map}. The obtained mass is M$_{\rm LTE} \sim 3.5 \times 10^{4}$ \msun.

\begin{figure}
\centering
\includegraphics[width=7cm]{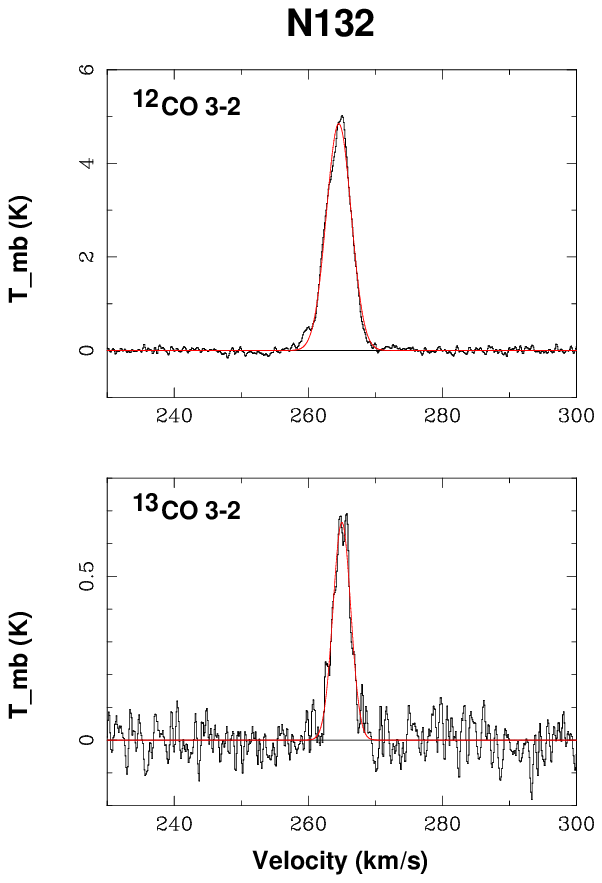}
\caption{Spectra of the CO isotopes detected towards N132. The Gaussian fitting to each
spectrum is shown in red.}
\label{spectraN132}
\end{figure}

\begin{figure}
\centering
\includegraphics[width=9cm]{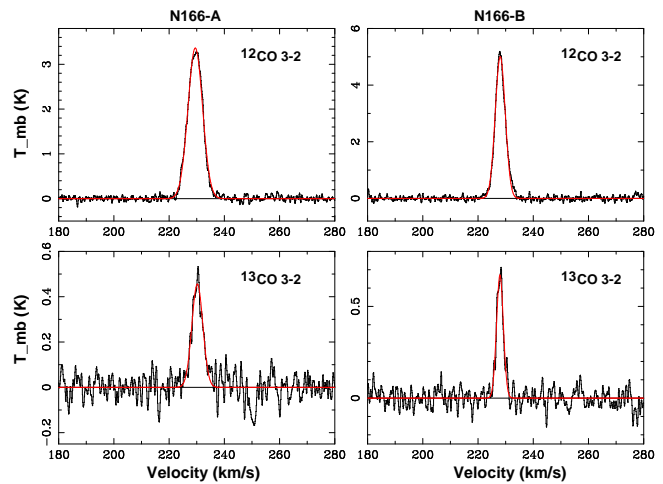}
\caption{Spectra of the CO isotopes detected towards both positions in N166. The Gaussian fitting to each
spectrum is shown in red.}
\label{spectraN166}
\end{figure}

Figure\,\ref{spectraN159} shows the spectra of the molecular lines observed towards N159-W (red cross in Fig.\,\ref{n159map}). 
This position is about 4\s~close to the location of the MYSO 053937.56-694525.4 catalogued in \citet{chen10} which is
very likely the responsible of the molecular outflows detected by \citet{fukui15}. Despite the high noise in the C$^{18}$O J=3--2 line
(about 25 mK), we included the spectrum as the signal is still quite evident.
Figures\,\ref{spectraN132} and\,\ref{spectraN166} show the CO isotopes spectra observed towards N132 and N166.
The line parameters from these spectra are presented in Table\,\ref{lines}. The peak main-beam temperature,
the central velocity, and the FWHM line width (Cols.\,3--5) were obtained from Gaussians fits (red curves
in the spectra figures).
Col.\,6 lists the integrated line intensity.
The C$_{2}$H J=4--3 line presents two peaks due to its fine structure components. One peak should correspond to the
blended C$_{2}$H (4--3) {\it J}=9/2--7/2 F=5--4 and 4--3 lines, and the other to the blended (4--3) {\it J}=7/2--5/2 F=4--3 and 3--2 lines
(see Paper\,I where it is presented the same detection towards N133, and the NIST data
base\footnote{http://www.nist.gov/pml/data/micro/index.cfm}). The HCN J=4--3 emission was fitted with two Gaussians, probably due to
a fine structure component, however, in Table\,\ref{lines} the integrated line intensity refers to the entire line.

\begin{table*}
\caption{Line parameters for the observed molecular lines.}
\label{lines}
\begin{tabular}{lccccc}
\hline
\hline
Molecular line    & Frequency        & T$_{\rm mb}$ peak & v$_{\rm LSR}$    & $\Delta$v (FWHM)  & $\int{\rm T_{mb} dv}$  \\
                  &  (GHz)           &   (K)             &   (\ks)          &  (\ks)            &           (K \ks)      \\ 
\hline 
\multicolumn{6}{c}{ {\bf N159}  }\\
\hline
\noalign{\smallskip}
\2 (3--2)        &   345.795       &   9.83              &  239.05         &  7.85             & 81.7$\pm$1.6           \\
\noalign{\smallskip}
\3 (3--2)        &   330.587     &   1.86              &  238.85         &  6.40             & 12.7$\pm$1.5           \\
\noalign{\smallskip}
C$^{18}$O (3--2)  &  329.330      &   0.05{\bf *}             &  235.68         &  4.73      & 0.47$\pm$0.25                   \\
\noalign{\smallskip}
CS (7--6)        &  342.882      &   0.11              &  237.89         &  3.06             & 0.37$\pm$0.05            \\
\noalign{\medskip}
\multirow{3}{*}{C$_{2}$H (4--3)  ~~$\left\lbrace \rule{0pt}{0.5cm} \right.$ } &    &   0.082  &    & 5.66  &   0.51$\pm$0.08\\
                    &   349.399            &                     &  237.59         &                   &                   \\
                    &               &   0.072             &                 &  6.23             & 0.49$\pm$0.07\\  
\noalign{\medskip}
HCO$^{+}$ (4--3)  & 356.734       &   1.03              &  238.47         &  6.30             & 7.0$\pm$1.2          \\
\noalign{\medskip}
\multirow{3}{*}{HCN (4--3) ~~$\left\lbrace \rule{0pt}{0.5cm} \right.$ } &     & 0.095    &  238.10         &  5.24   &   \\
                     & 354.505    &        &               &                   & 0.80$\pm$0.10          \\
                     &     & 0.035                &  232.70         &  7.00             &                        \\ 
\noalign{\medskip}    
HNC (4--3)           & 362.630    & 0.11                &  238.53         &  5.42             & 0.64$\pm$0.07          \\
\hline
\multicolumn{6}{c}{ {\bf N132}  }\\
\hline
\2 (3--2)            & 345.795    &   4.85              &  264.55         &  4.35             & 22.7$\pm$1.2            \\
\noalign{\smallskip}
\3 (3--2)            & 330.587    &   0.66              &  265.00         &  3.10             & 2.2$\pm$0.5             \\
\hline
\multicolumn{6}{c}{ {\bf N166-A}  }\\
\hline
\2 (3--2)      & 345.795          &   3.36              &  229.47         &  6.23             & 22.2$\pm$1.4             \\
\noalign{\smallskip}
\3 (3--2)      & 330.587          &   0.45              &  230.22         &  4.68             & 2.3$\pm$0.4              \\
\hline
\multicolumn{6}{c}{ {\bf N166-B}  }\\
\hline
\2 (3--2)     & 345.795           &   5.00              &  228.05         &  4.26             & 23.4$\pm$1.3           \\
\noalign{\smallskip}
\3 (3--2)     & 330.587           &   0.67              &  227.93         &  2.89             & 2.1$\pm$0.4            \\
%Llave:  $\left\lbrace \rule{0pt}{0.5cm}$
\hline
\multicolumn{6}{l}{\footnotesize {\bf *} The rms noise is 0.025 K.} \\
\end{tabular}
\end{table*}

\begin{table*}
\caption{Integrated intensity ratios.}
\label{ratios}
\begin{tabular}{lccccccc}
\hline
\hline
Ratio           &  N159$^{a}$     & N159$_{(1-0)}^{b}$  &  N113$^{c}$ &  N113$_{(1-0)}^{b}$ & N132$^{a}$   & N166-A$^{a}$   & N166-B$^{a}$    \\
\hline
$\frac{\rm ^{12}CO}{\rm ^{13}CO}$     &  $6.4\pm0.7$   &   9.12              &  $6.9\pm1.1$  &  7.28 &  $10.31\pm2.40$ & $9.65\pm1.02$ &  $11.14\pm1.11$ \\
\noalign{\medskip}
$\frac{\rm HCO^+}{\rm HCN}$           &  $8.7\pm1.8$   &  1.36              &  $4.8\pm1.1$   &  1.35 & --  & -- & --  \\
\noalign{\medskip}
$\frac{\rm HCN}{\rm HNC}$             &  $1.25\pm0.18$ &  3.90               &  $2.0\pm0.7$  &  2.82 & --  &  -- & --    \\
\noalign{\medskip}
$\frac{\rm HNC}{\rm HCO^{+}}$         &  $0.09\pm0.01$ &  0.18              &  $0.10\pm0.03$ &  0.26 & --  & --  & -- \\
\hline
\multicolumn{5}{l}{$^{a}$ This work.} \\
\multicolumn{5}{l}{$^{b}$ Ratios from the J=1--0 lines \citep{chin97}} \\
\multicolumn{5}{l}{$^{c}$ Ratios from the same lines used in this work (Paper\,I)} \\
\end{tabular}
\end{table*}

Table\,\ref{ratios} presents the integrated intensity ratios for some of the lines presented in Table\,\ref{lines}. For
comparison, the ratios obtained towards N159-W from the J=1--0 line by \citet{chin97} are also included. Table\,\ref{ratios} lists as well 
the ratios obtained towards N113 from Paper\,I.

\subsection{Non-LTE analysis of N159}
\label{nonlte}

\begin{table}
\caption{Radex results.}
\label{tradex}
\small
\centering
\begin{tabular}{lcc}
\hline
\hline
T$_{k}$ (K)            & n$_{\rm H_{2}}$ (cm$^{-3}$) & N(molec.) (cm$^{-2}$)  \\
\hline
\multicolumn{3}{c}{ {\bf \2}  }\\
\hline
20             &   $(1.1-1.8) \times 10^{4}$              &   $\sim 2.35 \times 10^{17}$   \\
80             &   $(0.1-1.1) \times 10^{6}$            &    $ (4.70-6.23) \times 10^{16}$     \\
\hline
\multicolumn{3}{c}{ {\bf \3}  }\\
\hline
20             &   $1.32 \times 10^{4}$            &    $\sim 1.00 \times 10^{16}$    \\
80             &     --                            &             --      \\
\hline
\multicolumn{3}{c}{ {\bf CS}  }\\
\hline
20             &   $4.68 \times 10^{6}$            &    $ 1.44 \times 10^{13}$    \\
80             &   $4.86 \times 10^{5}$            &    $ 1.28 \times 10^{13}$     \\
\hline
\multicolumn{3}{c}{{\bf HCO$^{+}$}}\\
\hline
20             &   $> 6.3 \times 10^{6}$            &    $ (8-9) \times 10^{12}$    \\
80             &   $6.76 \times 10^{5}$            &    $ 8.32 \times 10^{12}$    \\
\hline
\multicolumn{3}{c}{{\bf HCN}}\\
\hline
20             &   $5.20 \times 10^{6}$            &    $ 1.02 \times 10^{13}$    \\
80             &   $4.65 \times 10^{5}$            &    $ 7.74 \times 10^{12}$    \\
\hline
\multicolumn{3}{c}{{\bf HNC}}\\
\hline
20             &   $1.25 \times 10^{7}$            &    $ 2.23 \times 10^{12}$    \\
80             &   $3.02 \times 10^{6}$            &    $ 2.79 \times 10^{12}$    \\
\hline
\multicolumn{3}{l}{\footnotesize {\bf Note:} $\tau << 1$ in all cases for CS, HCO$^{+}$, HCN, and HNC.} \\
\multicolumn{3}{l}{\footnotesize $\tau \sim 1$ and $< 1$ for most of the cases in \2 and \3.} \\
\multicolumn{3}{l}{\footnotesize These results are obtained by assuming a beam filling factor of 0.5.}\\
\end{tabular}
\end{table}

\begin{figure}
\centering
\includegraphics[width=7.5cm]{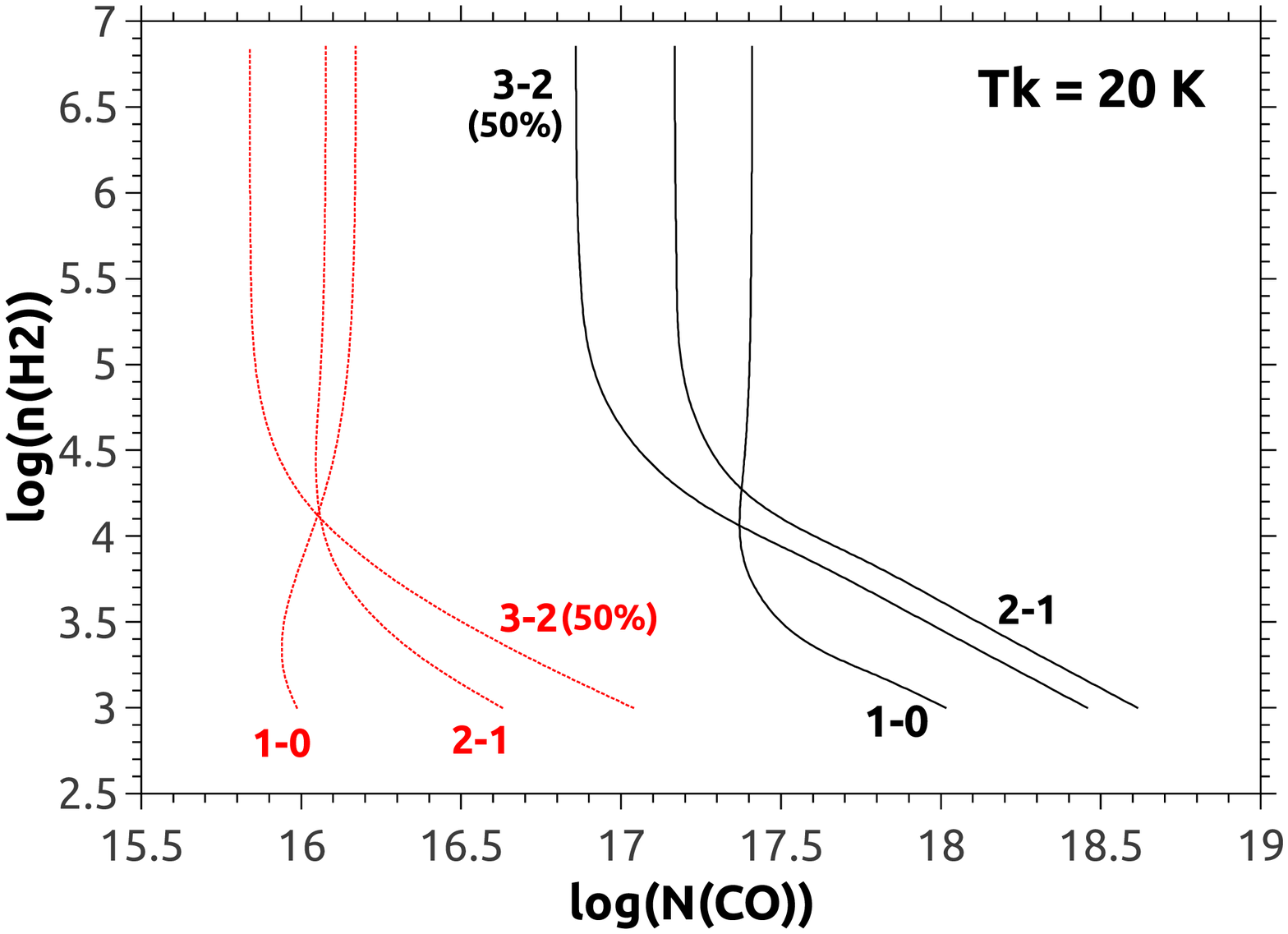}
\includegraphics[width=7.5cm]{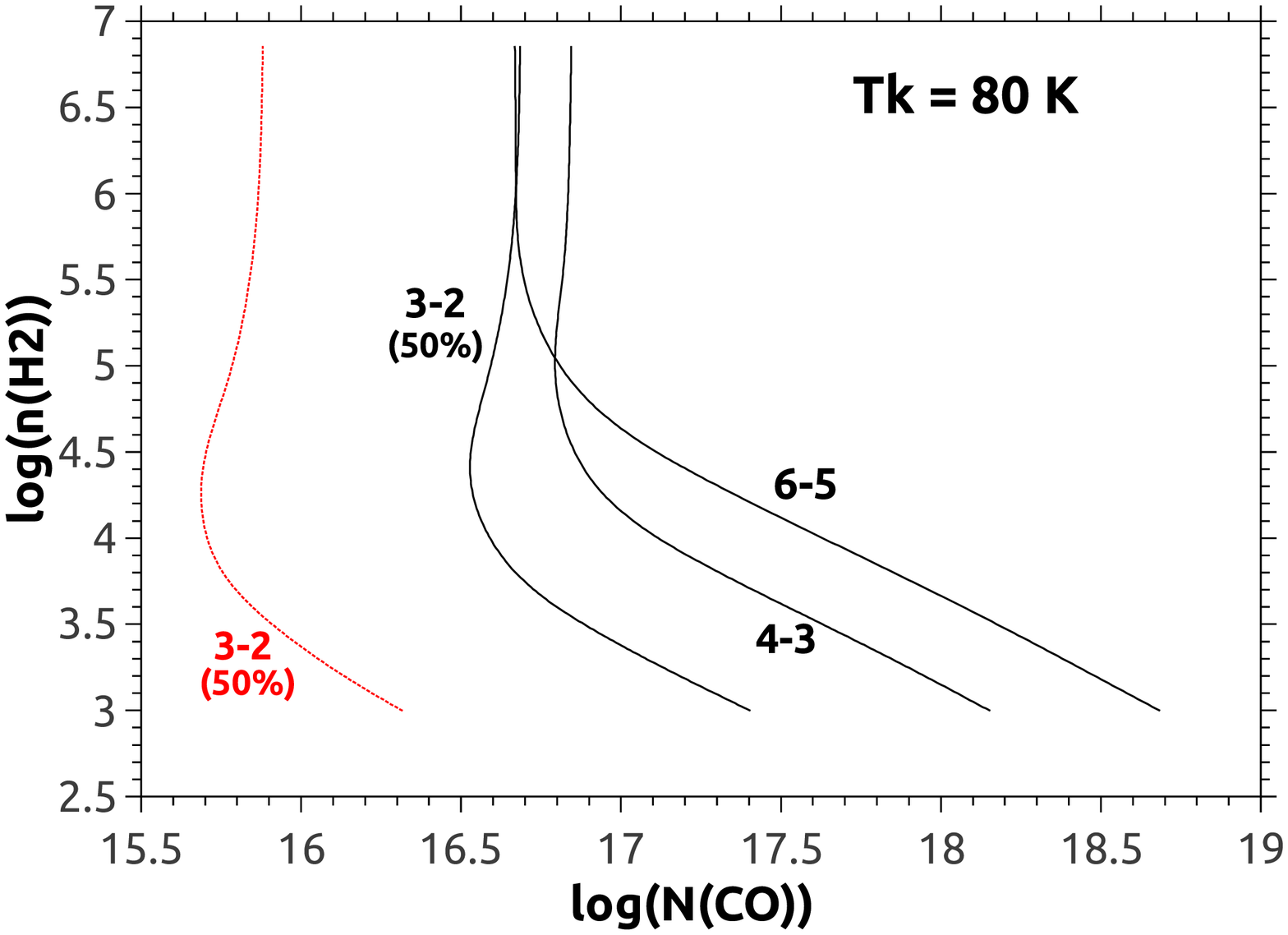}
\caption{Radex results for \2 (black lines) and \3 (red dashed lines). Data of \2 J=4--3 and 6--5 lines towards the same point of
our observations were kindly provided by Okada Y. \citep{okada15}. The J=1--0 and 2--1 lines are from \citet{chin97} and \citet{joha94},
respectively.}
\label{radexCO}
\end{figure}

\begin{figure}
\centering
\includegraphics[width=8.3cm]{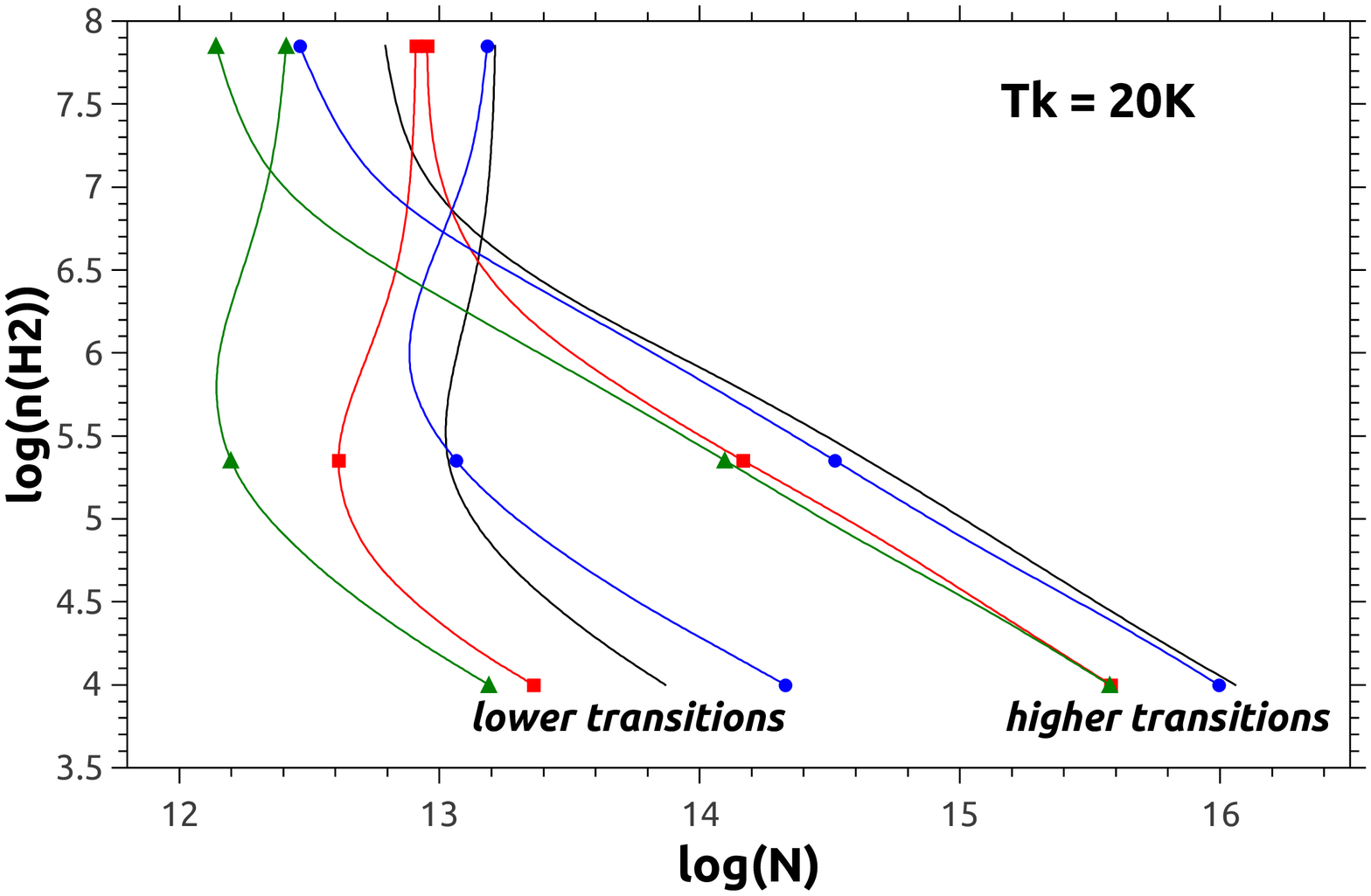}
\includegraphics[width=8.3cm]{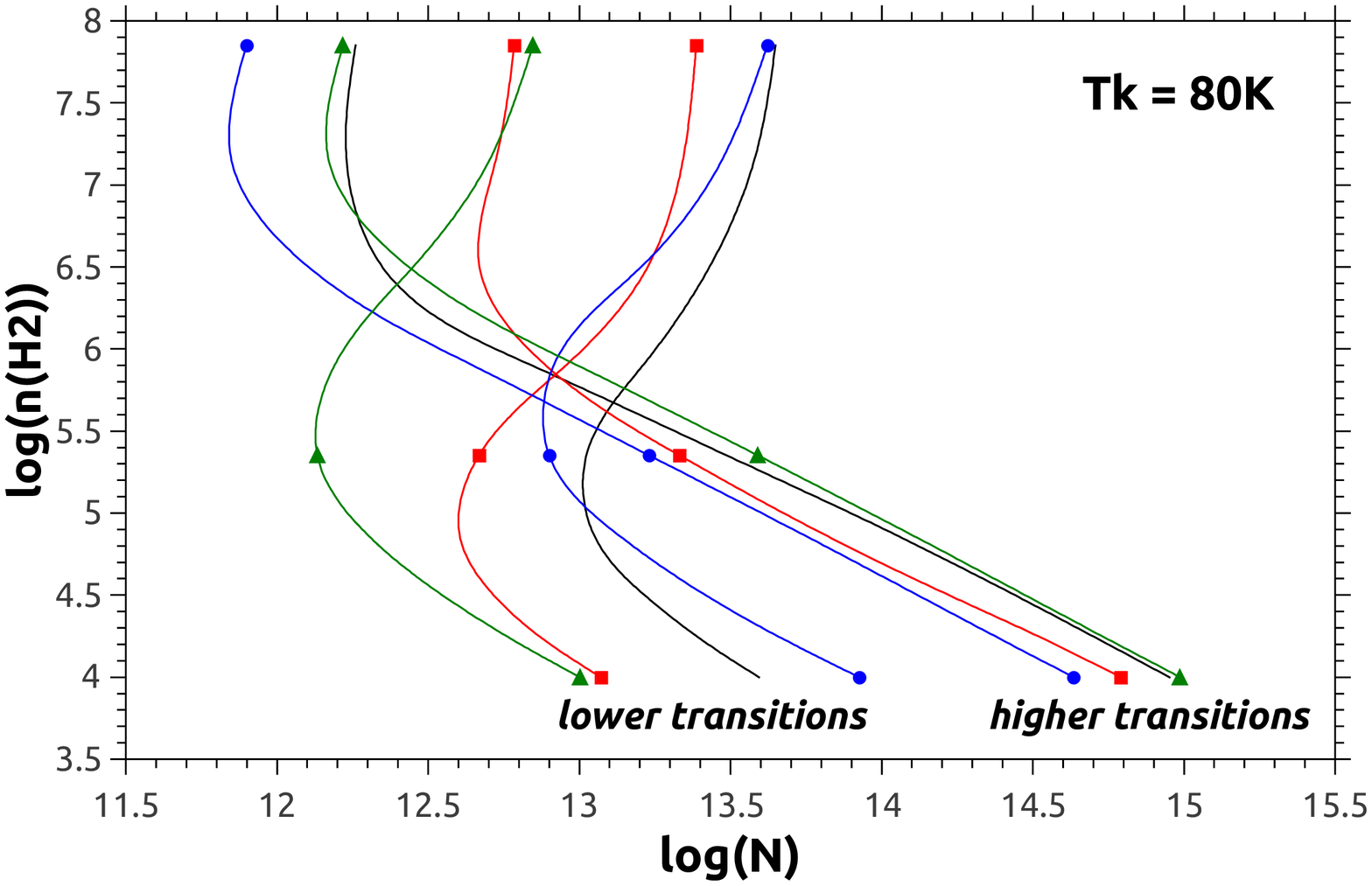}
\caption{Radex results for CS (black without symbols), HCN (red with squares), HNC (green with triangles), and \H~(blue with circles).}
\label{radex}
\end{figure}

Using the \2 and \3 J=1--0 line parameters from \citet{chin97}, the convolved \2 and \3 J=2--1 from \citet{joha94}, our results of 
the \2 and \3 J=3--2 lines, and the line parameters of \2 J=4--3 and J=6--5 towards the same position of our observation 
point at N159-W (T$_{\rm mb}^{4-3} = 8$ K, $\Delta$v$^{4-3} = 8.9$ \ks~and T$_{\rm mb}^{6-5} = 6.6$ K, $\Delta$v$^{6-5} = 7.6$ \ks; 
data kindly provided by Okada Y., see \citealt{okada15})  we
performed a non-LTE study of the CO using the RADEX code \citep{vander07}.
The main-beam temperatures were corrected for beam dilution by calculating T$_{\rm mb}^{\prime} =$ T$_{\rm mb}$/$\eta_{bf}$.
Following \citet{okada15}, we used a beam filling factor of $\eta_{bf} = 0.5$.
Figure\,\ref{radexCO}
presents the results obtained for kinetic temperatures of 20 and 80 K, displaying the expected H$_{2}$ density and the
column density pairs corresponding to a given T$_{\rm mb}^{\prime}$. The kinetic temperature values are due to
consider both the presence of a cold gas component (\citet{ott10} obtained a
T$_{\rm k} \sim 16$ K from ammonia lines), and a warmer one (likely due to the star forming processes and the radiation from massive
stars). In the warmer case the code was run for a grid of kinetic temperatures between 20 and 100 K. The selected model was that
in which the intersection of the curves is more tight (this was the model with T$_{\rm k} =$ 80 K). 
To perform this analysis, we assumed that the lower CO transitions arise mainly from the cold gas component, while
the higher ones from the warmer one. Given that it is likely that the J=3--2 transition arises from both components, a fifty percent 
of its emission was roughly assigned to each component.   

The same non-LTE analysis was done for the CS, \H, HCN, and HNC. The parameters for
the lowest transition of these molecular species were obtained from \citet{chin97} who observed a point 
located at $\sim$6\s~from our observation. The results are presented in Figure\,\ref{radex}. 
In the cold gas possibility the results for the \H~do not converge, and in addition, 
the intersections of all curves are not so tight as in the 80 K case.
Table\,\ref{tradex} presents the results for all analysed molecules.

\section{Discussion}

\subsection{N159}

Our \3 J=3--2 map of N159-W is very similar to the map recently presented by \citet{okada15}, which shows 
that the molecular peak is shifted southwest compared to the peak of the IR emission.
It is important to note that we are presenting the first detections of \H, HCN, HNC, and C$_{2}$H in the J=4--3 transition, CS in
the J=7--6, and tentatively the  C$^{18}$O J=3--2 line towards N159-W, precisely towards an IR peak, 
revealing that this region has the physical conditions
needed to excite these lines (for instance the critical densities of \H, HCN J=4--3, and CS J=7--6 are 1.8, 8.5, and 2.9 $\times 10^{6}$
cm$^{-3}$, respectively \citep{greve09}).
Concerning to the C$_{2}$H J=4--3 line, as done in N113 (Paper\,I),
we analyse the measured FWHM $\Delta$v of the peaks and
compare with the analysis presented in \citet{beuther08} towards a Galactic sample of star-forming regions in different evolutionary stages.
The authors showed that the C$_{2}$H J=4--3 lines towards ultracompact \hii~regions are significantly broader ($\Delta$v $=$ 5.5 \ks~in average) than
those obtained towards infrared dark clouds and high-mass protostellar objects, i.e. sources representing earlier stages in star formation.
Our C$_{2}$H $\Delta$v values are indeed broad, which is consistent with the position
of the molecular observation, almost in coincidence with a bright compact radio continuum source at R.A.$=$5:39:37.48, dec.$=-69$:45:26.10 (J2000)
\citep{hunt94,indeb04}, catalogued as a compact \hii~region likely generated by two O4V/O5V stars \citep{martin05}. Thus, we are
presenting more evidence supporting that the chemistry involving the C$_{2}$H in compact and/or UC\hii~regions in the LMC should be similar to
that in Galactic ones.

The comparison between the ratios from higher transition and those obtained from the lower one presented in Table\,\ref{ratios} 
shows the same trend for both N159 and N113, i.e. ratios from J=4--3 are lower than those from J=1--0 
except for $\frac{\rm HCO^+}{\rm HCN}$ in both regions. The discrepancy
between both ratios is more pronounced in N159 than in N133.
On the other side, the $\frac{\rm HCO^+}{\rm HCN}$ ratio in the J=4--3 line in N159 is almost twice the value derived in N113. This may suggest an
overabundance of HCO$^{+}$ in N159 that is not evidenced in the lower line ratio, probably due to line saturation effect. However,
as none excitation effects are taking into account in the line ratios, the overabundance statement is far to be conclusive.

The RADEX results suggest the presence of both cool and warm gas in the analysed region. Indeed, the CO emission at the observed 
position likely arises from both, gas at 20 K with a density about $1.5 \times 10^{4}$ cm$^{-3}$, and gas at 80 K 
with densities between $10^{5}$ and $10^{6}$ cm$^{-3}$. The CO column density in the warm gas component is $\sim5$ times 
lesser than in the colder one. 
Our results for the warm gas component are in close agreement with what was obtained by \citet{pineda08}, who used the
\2 and \3 J=7--6, 4--3 and 1--0 lines, and some lines of [CI] and [CII].
Additionally, the RADEX results for the CS, \H, HCN, and HNC indicate that it is very likely that their emissions arise
mainly from warm gas with densities between $5 \times 10^{5}$ to some $10^{6}$ cm$^{-3}$, which is in agreement with the CO warm results.

It is known that the $\frac{\rm HCN}{\rm HNC}$ abundance ratio depends on kinetic temperature \citep{schilke92}. 
From the obtained column densities we derived $\frac{\rm HCN}{\rm HNC} \sim 2.77$, which is compatible with warm gas (T$_{\rm k}$ between 
50 and 100 K) \citep{hel}. In addition, a $\frac{\rm HCN}{\rm HNC}$ ratio greater than 1 agrees with an star-forming scenario \citep{sch02},
and in particular, our value is very similar to the values obtained towards active cores in galactic infrared dark clouds
\citep{jin15}. The $\frac{\rm HCN}{\rm HNC}$ ratio greater than 1 can be explained by the rapid C $+$ HNC $\rightarrow$ C $+$ HCN reaction that works
as long as the carbon atom abundance is still high \citep{loison14}, which seems to be the case, since according to \citet{okada15}
N159-W has the highest C column density within the N159 complex.
This ratio could confirm, in an independent way, the existence of warm gas in the studied region. However, we should be cautious with such ratios 
because the obtained column densities are dependent on the assumed beam filling factor.

\subsection{N132 and N166}

In the case of N132 and N166 only \2 and \3 were detected. The non-detection of higher density gas tracers such as C$^{18}$O,
N$_{2}$H$^{+}$, and DCO$^{+}$ (which were observed in our observation run with integration times of 560 and 1440 sec) 
may indicate that the density of the molecular gas in these regions 
is not so high. This is in agreement with what \citet{mina08} concluded for N166 and is consistent with the higher \2/\3 integrated
intensity ratio compared with the denser regions N159 and N113 (see Table\,\ref{ratios}).
The ratios obtained in N166 are in good agreement with the ratios presented in \citet{garay02} using the J=1--0
line for several clouds of the giant molecular Complex-37. N166-A and -B are about 19\s~and 27\s~close
to N166-Clump 2 and N166-Clump 1 belonging to Complex-37 \citep{mina08}. From an LVG analysis the authors point out that the studied clumps
in N166 have a density between some 10$^{2}$ to a few 10$^{3}$ cm$^{-3}$ with kinetic temperatures between 25 and 150 K.
By assuming LTE we obtain similar values of T$_{\rm ex}$ (between 24 and 30 K) and \2 and \3 
optical depths ($\tau_{12} \sim 7$ and $\tau_{13} \sim  0.14$)
for N132 and N166, suggesting that the physical conditions should be similar in both regions.

\section{Summary}

We have performed a molecular line study towards the LMC \hii~regions N159, N132, and N166 in the 345 GHz window 
using ASTE. We mapped a 2\farcm4 $\times$ 2\farcm4  region towards the
molecular cloud N159-W in the \3 J=3--2 line and several molecular lines as single pointings at an IR peak,
about 4\s~close to the position of a MYSO.
In addition, several molecular lines were also observed towards two positions in N166 and
one position in N132, resulting in positive detections only the \2 and \3 J=3--2.  
Our main results can be summarized as follows:

(a) Our \3 J=3--2 map of N159-W is very similar to the map recently presented by \citet{okada15} and shows 
that the molecular peak is shifted southwest compared to the peak of the IR emission.
We estimated the LTE mass of the molecular clump in $\sim3.5 \times 10^{4}$ \msun.

(b) Towards the IR peak position of N159-W we detected emission from HCN, HNC, \H, C$_{2}$H J=4--3, CS J=7--6,
and tentatively C$^{18}$O J=3--2, being the first reported detection of these molecular lines in this region. In addition
it was obtained an spectrum of \2 and \3 J=3--2 towards this position. The detection of the mentioned molecular species
in the 345 GHz window proves the presence of high-density gas and shows the usefulness of performing surveys in this wavelength
window to increase our knowledge about the physical and chemical conditions of the ISM in the LMC.

(c) The detection and the line width of C$_{2}$H J=4--3 towards N159-W is compatible with an environment affected by the 
action of an \hii~region. Following 
our previous study in N113, we conclude that we are presenting more 
evidence supporting that the chemistry involving this molecular species in compact and/or UC\hii~regions in the LMC should be similar 
to that in Galactic ones.

(d) Using our observed CO lines and several lines of this molecule from the literature we performed a non-LTE study which suggests 
that the CO emission likely arises from both, gas at 20 K with a density about $1.5 \times 10^{4}$ cm$^{-3}$, and gas at 80 K
with densities between $10^{5}$ and $10^{6}$ cm$^{-3}$.
The same non-LTE analysis for the CS, \H, HCN, and HNC shows that we are indeed probing high-density gas ($5 \times 10^{5}$ to some $10^{6}$ cm$^{-3}$) 
and it is very likely that their emissions arise mainly from warm gas, which is in agreement with the CO warm results.

(e) Using the column densities derived from the non-LTE study we obtained a $\frac{\rm HCN}{\rm HNC}$ abundance ratio greater than 1,
which is compatible with warm gas and with an star-forming scenario. This is
in agreement with the presence of MYSOs in the studied region, one of them driving molecular outflows.

(f) Based on the CO line analysis and the non-detection of higher density tracers we suggest that N132 and N166 should have similar 
physical conditions, with densities between some 10$^{2}$ to a few 10$^{3}$ cm$^{-3}$ and kinetic temperatures between 25 and 150 K.

\section*{Acknowledgments}

We acknowledge the anonymous referee for her/his helpful comments and suggestions.
We wish to thank to Y. Okada for kindly provide us with the CO higher transitions data.
Thanks to Bastiaan Zinsmeister for his contribution to this paper. 
The ASTE project is led by Nobeyama Radio Observatory (NRO), a branch
of National Astronomical Observatory of Japan (NAOJ), in collaboration
with University of Chile, and Japanese institutes including University of
Tokyo, Nagoya University, Osaka Prefecture University, Ibaraki University,
Hokkaido University, and the Joetsu University of Education.
S.P. and M.O. are members of the {\sl Carrera del 
investigador cient\'\i fico} of CONICET, Argentina. 
This work was partially supported by grants awarded by CONICET, ANPCYT and UBA (UBACyT) from Argentina.
M.R. wishes to acknowledge support from FONDECYT(CHILE) grant N$^{\rm o}$1140839.

\label{lastpage}

\end{document}